\documentclass[10pt]{article}
\usepackage{hhline}

\begin{document}

\newcommand{\beq}{\begin{equation}}
\newcommand{\eeq}{\end{equation}}
\newcommand{\beqn}{\begin{eqnarray}}
\newcommand{\eeqn}{\end{eqnarray}}
\newcommand{\bsig}{\mbox{\boldmath{$\sigma$}}}
\newcommand{\btau}{\mbox{\boldmath{$\tau$}}}
\newcommand{\bnab}{\mbox{\boldmath{$\nabla$}}}
\newcommand{\br}{\mbox{${\bf r}$}}
\newcommand{\bR}{\mbox{${\bf R}$}}
\newcommand{\bp}{\mbox{${\bf p}$}}
\newcommand{\bn}{\mbox{${\bf n}$}}
\newcommand{\ep}{\mbox{\varepsilon}}

\title{ \Large \bf On shielding of nuclear electric dipole moments in atoms}

\author{\\ V.F. \ Dmitriev$^1$, I.B. \ Khriplovich$^1$, and R.A. \ Sen'kov$^{1,2}$\\
{\small $^1$ Budker Institute of Nuclear Physics,} \\ {\small
Lavrentjev pr. 11, Novosibirsk, 630090, Russia} \\
{\small $^2$ Department of Physics, Novosibirsk State
University,}\\{\small  Pirogov st. 2, Novosibirsk, 630090,
Russia}}

\maketitle

\begin{abstract}

{\normalsize

We demonstrate explicitly that some recent calculations of atomic
electric dipole moments (EDM) are incomplete. A contribution
overlooked therein is pointed out. When included, it cancels
exactly the result of those calculations, and thus restores the
standard conclusions for nuclear EDM in atoms.}

\end{abstract}

\vspace{1cm}

The existence of EDMs of elementary particles, nuclei, and atoms
is forbidden by CP invariance. The predictions of the standard
model of electroweak interactions for these EDMs are at least six
orders of magnitude below the present experimental bounds. It
makes experimental searches for EDMs, even at present level of
accuracy, extremely sensitive to possible new physics beyond the
standard model.

The best upper limit on EDM of anything was obtained on the
$^{199}$Hg atomic dipole moment~\cite{rgf01}:
\begin{equation} \label{eq:intro-exp}
d(^{199}{\rm Hg}) < 2.1 \times 10^{-28}\; e\mbox{ cm}.
\end{equation}

Unfortunately, the implications of this result for the dipole
moment of the valence neutron of the $^{199}$Hg nucleus is much
less impressive~\cite{our1}:
\beq\label{eq:intro-1}
|d_n| < 4.0 \times 10^{-25} \; e\mbox{ cm}
\eeq
(still, it is not so far away from the results of the direct
measurements of the neutron EDM~\cite{n1,n2}: $|d_n| < (0.6 - 1.0)
\times 10^{-25} \; e$ cm).

The explanation is as follows. For a neutral atom of point
particles, in equilibrium under the action of electrostatic
forces, there is no effect due to the EDMs of the constituent
particles~\cite{pr50,gl}. Indeed, the atom remains at rest when an
external electric field ${\bf E}_{ext}$ is applied. This comes
about by an internal rearrangement of the system's constituents
giving rise to an internal field ${\bf E}_{int}$ that exactly
cancels ${\bf E}_{ext}$ at each charged particle, as required by
the static equilibrium condition. Thus, there is no observable
effect due to dipole moments of the system's constituents; the
external field is effectively switched off.

A quantum mechanical proof of this statement was given
in~\cite{sch63}. It was also pointed out therein that the
shielding of atomic nucleus is not complete, observable nuclear
EDM effects arise when one takes into account its finite size. Of
course, the resulting suppression of these effects is quite
essential. The suppression factor $\eta$ is roughly the ratio
squared of the nuclear radius to that of the atomic K-shell (see,
for instance,~\cite{kl}):
\beq\label{sch}
\eta \sim \left ( \frac{A^{1/3}r_0}{a/Z} \right)^2
\eeq
(here $a=0.5\times 10^{-8}$ cm, $r_0 =1.2\times 10^{-13}$ cm).
Numerically for the mercury atom  $\eta \sim 10^{-4}$, and this is
the reason why the strict atomic upper limit (\ref{eq:intro-exp})
reduces to much milder neutron one (\ref{eq:intro-1}).

However, this conclusion was revised in recent papers \cite{fuj,
fuj1}. The relations between atomic and neutron EDMs advocated
therein are as follows:
\begin{eqnarray} \label{eq:rez1}d(^{199}{\rm Hg}) \simeq -2.8 d_n , \\
\label{eq:rez2} d(^{129}{\rm Xe}) \simeq  1.6 d_n ,\\
\label{eq:rez3} d({\rm D}) \simeq 0.017 d_n,
\end{eqnarray}
where $d(^{129}{\rm Xe})$, $ d(^{199}{\rm Hg})$, and $d({\rm D})$
are the EDMs of xenon, mercury, and deuterium atoms, respectively.
In particular, from the upper limit (\ref{eq:intro-exp}) for the
dipole moment of mercury atom the authors of \cite{fuj} extract a
very strict bound on the neutron EDM
\beq\label{eq:intro-2}
d_n \simeq (0.37\pm 0.17 \pm 0.14)\times 10^{-28}\, e \mbox{ cm} .
\eeq

If results (\ref{eq:rez1}) -- (\ref{eq:intro-2}) were correct,
they would be, as discussed at length in~\cite{fuj1}, of paramount
importance for the problem of CP violation (as well as for the
present programs of searches for the neutron, nuclear and atomic
EDMs). This is why we believe that it is proper to analyze
attentively the arguments of~\cite{fuj}. To make our discussion as
simple and transparent as possible, we confine it to the case of
deuterium (where relation (\ref{eq:rez3}) differs from that
following from estimate (\ref{sch}) by 8 orders of magnitude). An
important contribution to the atomic EDM overlooked in~\cite{fuj}
is pointed out, which exactly cancels the contribution considered
therein. So, let us discuss in detail how the shielding works.

The unperturbed deuterium Hamiltonian is
\beq\label{h}
H =  { {{\bf p}}^2\over{2m} } -  {e^2\over{|{\bf r} - {\bf R}/2|}}
+{ {\bf P}^2\over{M} }+ U({\bf R},{\bf S}).
\eeq
Here ${\bf p}$ and ${\bf r}$ are the momentum and coordinate of
electron, ${\bf P}$ and ${\bf R}$ are the relative momentum and
coordinate of the nucleons, $U({\bf R},{\bf S})$ is the strong
proton-neutron potential (by the way, it depends essentially on
the total spin {\bf S} of the nucleons); all coordinates are
counted off the center of mass of the deuteron. We confine here
and below to the nonrelativistic limit, which is quite sufficient
for our purpose.

To separate the atomic and nuclear variables, we rewrite
Hamiltonian~(\ref{h}) as $H=H_a + H_N + W$, where
\beq\label{ha}
H_a =  { {{\bf p}}^2\over{2m} } -  {e^2\over r}
\eeq
is the atomic Hamiltonian,
\beq\label{hN}
H_N = { {\bf P}^2\over{M} }+ U({\bf R},{\bf S}).
\eeq
is the nuclear one; perturbation
\beq\label{w}
W= - {e^2\over{|{\bf r} - {\bf R}/2|}}\,+\,{e^2\over{r}}\,=
\,\frac{\bR}{2}\,\bnab\,\frac{e^2}{r}\,=-\,\frac{\bR}{2}\,
i\left[\bp, H_a \right]
\eeq
is treated to first order in $R/r$ only.

The next perturbation describes the interaction of the electron
and proton charges with an external electric field ${\bf
E}_{ext}$:
\beq\label{V}
V=\,e\,\left(\br - \,\frac{\bR}{2}\right){\bf E}_{ext}\,;
\eeq
here and below $e=e_p>0$.

And at last, we present the P odd and T odd interaction of the
neutron EDM ${\bf d}_n$ with the Coulomb field of the proton
considered in~\cite{fuj}:
\beq\label{dnp}
v\,= \,e\,{{\bf d}_n  {\bf R}\over{R^3}}\,.
\eeq
When combined with perturbation (\ref{V}), it results in the
following second-order contribution to the deuteron EDM:
\beq\label{d2}
{\bf d}_2 = \sum_{n} \;{  \langle 0 |e {\bf R}/2 |n \rangle
\langle n | v | 0 \rangle \over{E_0-E_n} } + \mbox{h.c.}\;.
\eeq
Of course, only that part of perturbation (\ref{V})
\beq\label{V1}
V_1=\,-\,e \,\frac{\bR}{2}\,{\bf E}_{ext}\,,
\eeq
which depends on the nuclear coordinate ${\bf R}$, is operative
here. Expression (\ref{d2}) for the nuclear EDM, induced by
interaction (\ref{dnp}), is certainly correct. Moreover, it is
valid for any P odd and T odd proton--neutron interaction, not
only for that described by formula (\ref{dnp}). In particular, the
deuteron EDM induced in this way by the P odd and T odd pion
exchange, was calculated in~\cite{kk}.

However, this is the EDM of the nucleus, deuteron, but not the
total EDM of the atom, deuterium, which should vanish in the point
limit due to the atomic shielding effect pointed out above.

To restore this shielding, we switch on, in line with (\ref{V})
and (\ref{dnp}), perturbation (\ref{w}). Its matrix element
between two atomic states with energies
$\varepsilon_2,\;\varepsilon_1$ can be conveniently rewritten as
follows:
\beq\label{w1}
\langle \varepsilon_2|\,W\,|\varepsilon_1 \rangle =
\,\frac{i}{2}\,\bR \langle \varepsilon_2 |\,\bp\,|\varepsilon_1
\rangle (\varepsilon_2-\varepsilon_1).
\eeq
As to interaction (\ref{V}), it obviously reduces in this case to
\beq\label{V2}
V_2=\,e\,\br {\bf E}_{ext}.
\eeq
Now we calculate the combined result of interactions (\ref{dnp}),
(\ref{w1}), and (\ref{V2}). After some rearrangement of terms,
using extensively the completeness relation for atomic states, we
obtain the following result for this third-order contribution:
\beq\label{d3}
{\bf d}_3 = - \sum_{n} \;{  \langle 0 | e {\bf R}/2 |n \rangle
\langle n | v | 0 \rangle \over{E_0-E_n} } + \mbox{h.c.}\,.
\eeq

The contributions (\ref{d2}) and (\ref{d3}) cancel, in complete
accordance with the shielding theorem. Obviously, the cancellation
occurs for any P odd and T odd proton-neutron interaction $v$.

One should not be surprised by the cancellation between
second-order effect (\ref{d2}) and third-order one (\ref{d3}).
While perturbation (\ref{dnp}) is common for both effects, it can
be easily checked that the combined action of perturbations
(\ref{w1}) and (\ref{V2}) can well be on the same order of
magnitude as that of (\ref{V1}).

In fact, the neutron and proton electric dipole moments, ${\bf
d}_n$ and ${\bf d}_p$, induce the EDM of the atom due to the
so-called Schiff moment (SM)~[2, 12--14] (SM vanishes, of course,
in the limit of point nucleus). The general expression for SM
operator, induced by ${\bf d}_n$ and ${\bf d}_p$, reduces for the
deuteron to the following form:
\beq\label{sm} {S}_i = \,\frac{1}{24}\, ({d_n}_i + {d_p}_i)\,(R^2 -
\langle R^2\rangle) + \,\frac{1}{60}\, ({d_n}_j + {d_p}_j)\,[\, (
3 R_i R_j - \delta_{ij} R^2)  - 4 Q_{ij}\,]\,;
\eeq
here $\langle R^2\rangle$ and $Q_{ij}$ are the expectation values
of $R^2$ and the deuteron quadrupole moment, respectively.

In conclusion, the following peculiarity of the deuteron is worth
mentioning. If the strong proton-neutron potential $U({\bf R},{\bf
S})$ were independent of the total spin {\bf S}, we would have,
obviously,
\[
\langle({d_n}_i + {d_p}_i)\,R^2\,\rangle = \langle\,{d_n}_i +
{d_p}_i\,\rangle\,\langle R^2\rangle, \;\; \;{\rm and}\;\;\;
\langle({d_n}_j + {d_p}_j)\,( 3 R_i R_j - \delta_{ij}
R^2)\,\rangle =\,4\langle\,{d_n}_j + {d_p}_j\,\rangle\,Q_{ij}\;.
\]
Therefore, in this case the expectation value of the deuteron
Schiff moment (\ref{sm}), generated by ${\bf d}_n$ and ${\bf
d}_p$, would vanish, and together with it the atomic deuterium EDM
would vanish as well.

\begin{center} *** \end{center}

We are grateful to N. Auerbach and S. Lamoreaux for bringing
Refs.~\cite{fuj,fuj1} to our attention.

\end{document}